\documentclass[11pt]{article}
\usepackage{epsfig}

\newcommand{\BABARPubYear}    {04}

\newcommand{\BABARConfNumber} {005}
\newcommand{\SLACPubNumber} {10602}

\input babarsym

\providecommand{\btodspi}{\mbox{$B^0\to D_{s}^+\pi^-$}}

\providecommand{\btodsspi}{\mbox{$B^0\to D_{s}^{*+}\pi^-$}}

\providecommand{\btodssk}{\mbox{$B^0\to D_{s}^{*-}K^+$}}
\providecommand{\btodsospi}{\mbox{$B^0\to D_{s}^{(*)+}\pi^-$}}
\providecommand{\btodsrho}{\mbox{$B^0\to D_{s}^+\rho^-$}}

\providecommand{\btodssrho}{\mbox{$B^0\to D_{s}^{*+}\rho^-$}}

\providecommand{\btodrho}{\mbox{$B^0\to D^-\rho^+$}}

\providecommand{\btodrhocab}{\mbox{$B^0\to D^+\rho^-$}}

\providecommand{\btodskstar}{\mbox{$B^0\to D_{s}^{-} K^{*+}$}}
\providecommand{\btodsa}{\mbox{$B^0\to D_{s}^{+} a_{1}^{-}$}}
\providecommand{\btodssa}{\mbox{$B^0\to D_{s}^{*+} a_{1}^{-}$}}

\providecommand{\bchtodsrho}{\mbox{$B^+\to D^+_s\rho^0$}}
\providecommand{\bchtodssrho}{\mbox{$B^+\to D^{*+}_s\rho^0$}}

\providecommand{\dsphipi}{\mbox{$D^+_{s}\to \phi\pi^+$}}

\providecommand{\dsksk}{\mbox{$D^+_{s}\to \Kzb K^+$}}
\providecommand{\dskstark}{\mbox{$D^+_{s}\to \Kstarzb K^+$}}

\providecommand{\ds}{\ensuremath{D_{s}^+}}

\providecommand{\De}{\ensuremath{\Delta E}}
\providecommand{\mse}{\ensuremath{m_{ES}}}

\providecommand{\stwobg}{\ensuremath{\sin(2\beta+\gamma)}\xspace}

\providecommand{\brdsrholim}{\BR(\btodsrho)\ensuremath{<1.9\times10^{-5}}}
\providecommand{\lambdalim}{\ensuremath{r({D\rho})<9.5\times 10^{-3}}}
\providecommand{\Lumi}{$90\times10^6$}

\long\def\inst#1{\par\nobreak\kern 4pt\nobreak
    {\it #1}\par\vskip 10pt plus 3pt minus 3pt}

\begin{document}
{\pagestyle{empty}

\begin{flushright}
\babar-CONF-\BABARPubYear/\BABARConfNumber \\
SLAC-PUB-\SLACPubNumber \\
August 2004 \\
\end{flushright}

\par\vskip 4cm

% Title of the paper
\begin{center}
\Large \bf A Search for the Rare Decay \btodsrho
\end{center}

\bigskip

\begin{center}
\large The \babar\ Collaboration\\
\mbox{ }\\
\today
\end{center}
\bigskip \bigskip

% Abstract
\begin{center}
\large \bf Abstract
\end{center}
We report on a search for the decay \btodsrho\  
in a sample of \Lumi\ \FourS\ decays into $B$ meson pairs collected between 1999 and 2001 with the \babar\ detector
at the \pep2\ asymmetric-energy $e^{+}e^{-}$ collider. 
No significant excess of signal events above the expected background
is observed. We set a 90\% C.L. limit on the branching fraction \brdsrholim. Assuming a flavor
SU(3) symmetry relation between the decays \btodsrho\ and \btodrhocab,
we set a limit on the ratio of CKM-suppressed to CKM-favored
amplitudes \lambdalim\ at 90\% C.L. All results are preliminary.

\vfill
\begin{center}

Submitted to the 32$^{\rm nd}$ International Conference on High-Energy Physics, ICHEP 04,\\
16 August---22 August 2004, Beijing, China

\end{center}

\vspace{1.0cm}
\begin{center}
{\em Stanford Linear Accelerator Center, Stanford University, 
Stanford, CA 94309} \\ \vspace{0.1cm}\hrule\vspace{0.1cm}
Work supported in part by Department of Energy contract DE-AC03-76SF00515.
\end{center}

\newpage
} % end of pagestyle{empty}

% Input author list file
\begin{center}
\small

The \babar\ Collaboration,
\bigskip

%% author list as of 02-Jul-2004 (609 authors)
%
B.~Aubert,
R.~Barate,
D.~Boutigny,
F.~Couderc,
J.-M.~Gaillard,
A.~Hicheur,
Y.~Karyotakis,
J.~P.~Lees,
V.~Tisserand,
A.~Zghiche
\inst{Laboratoire de Physique des Particules, F-74941 Annecy-le-Vieux, France }
% ...

\end{center}\newpage

% The body of the paper starts here
\section{Introduction}

The Cabibbo-Kobayashi-Maskawa (CKM) quark flavor-mixing
matrix~\cite{CKM} provides an elegant explanation of the origin of
\CP\ violation within the Standard Model. 
\CP\ violation manifests itself as a non-zero area of the unitarity triangle~\cite{Jarlskog}.
While it is sufficient to measure one of the angles to demonstrate the
existence of \CP\ violation,
the unitarity triangle needs to be over-constrained by experimental
measurements 
in order to demonstrate that the CKM mechanism is the correct 
explanation of this phenomenon. One of the important measurements is
constraining the angle 
$\gamma = {\rm arg}(-V_{ud}V_{ub}^{*}/V_{cd}V_{cb}^{*})$ 
of the unitarity triangle. A
measurement of 
\stwobg\ can be obtained from the study 
of the time evolution of the $\Bz{\to} D^{(*)-} \pi^+$~
and $\Bz{\to} D^{(*)-} \rho^+$~\cite{chconj} decays, a large sample
of which is already available at the B-factories, and of the corresponding
CKM suppressed modes $\Bz{\to} D^{(*)+}\pi^-$ and 
$\Bz{\to}D^{(*)+}\rho^-$~\cite{sin2bg}. The first measurements of
\stwobg\ using $\Bz{\to} D^{(*)\mp}\pi^\pm$ decays have been recently
published~\cite{ref:s2bgDPi,ref:s2bgDRho}, and a
similar analysis using   
$\Bz{\to} D^{\mp}\rho^\pm$ decays is being reported at this
conference~\cite{ref:s2bgDRho}. 

%This measurement 
The measurement of \stwobg\ using $\Bz{\to} D^{\mp}\rho^\pm$ decays 
requires the knowledge of the ratio of the decay amplitudes, 
$r({D\rho})=|A(\Bz{\to} D^{+}\rho^-)/A(\Bz{\to}D^{-}\rho^+)|$. 
Unfortunately, the direct measurement of the branching fraction
$\BR(\Bz{\to} D^{+}\rho^-)$ is not 
possible with the currently available data sample due to the presence
of the copious background from $\Bzb{\to} D^{+}\rho^-$.
However, assuming SU(3) flavor symmetry, $r({D\rho})$ can be related
to the branching fraction of the decay \btodsrho~\cite{sin2bg}:
\begin{equation}
r({D\rho}) = 
  (\tan\theta_c)\frac{f_D}{f_{D_s}}\sqrt{\frac{\BR(\btodsrho)}{\BR(\btodrho)}}
\label{eq:rDRho}
\end{equation}
where $\theta_c$ is the Cabibbo angle, and $f_D/f_{D_s}$ is the ratio
of $D$ and $D_s$ meson decay constants~\cite{fdsd}. Other
SU(3)-breaking effects are typically assumed
to be of order $30\%$~\cite{ref:s2bgDPi}.

\begin{figure}[h]
\begin{center}
\epsfig{file=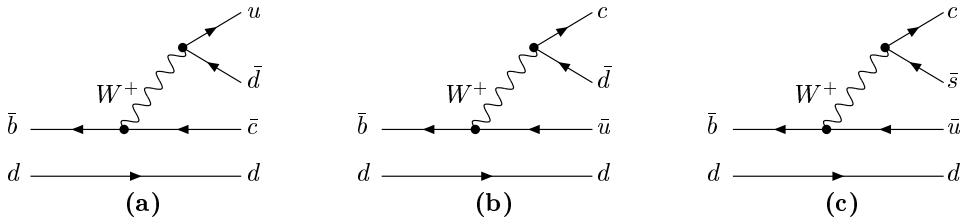,width=5in} 
\end{center}
\vspace{-0.5cm}
\caption{Dominant Feynman diagrams for CKM-favored decay
$\Bz\to D^-\rho^+$ (a), doubly CKM-suppressed decay
$\Bz\to D^+\rho^-$ (b), and the SU(3) flavor symmetry related decay 
\btodsrho\ (c).}   
\label{fig:diag} 
\end{figure}

Since $\Ds\rho^-$ has four
different quark flavors in the final state, only a single amplitude
contributes to the decay. The presence of the \Ds\ meson makes such
decays easy to identify. Fig.~\ref{fig:diag} 
shows the dominant Feynman diagrams for the decays 
$\Bz\to D^-\rho^+$, $\Bz\to D^+\rho^-$, and
\btodsrho. Eq.~(\ref{eq:rDRho}) assumes that 
the color-suppressed direct $W$-exchange amplitude for $\Bz\to
D^+\rho^-$ is negligibly small, which is supported by the
data~\cite{dspi}.  

The present limit on the branching fraction
$\BR(\btodsrho)$ is $7\times 10^{-4}$ at 90\% C.L.~\cite{PDG2004}. The
related decay $\btodspi$ has recently been observed~\cite{dspi}.
%\footnote[1]{From this point forward, charge-conjugation is assumed for all modes discussed.}

\section{Data Sample, Detector, and Simulation}
We use a sample of \Lumi\ \FourS\ decays into $B\overline{B}$
 pairs 
 collected in the years 1999-2002 with the \babar\ detector
at the \pep2\ asymmetric-energy $B$-factory~\cite{pep}.
Since the \babar\ detector is described in detail
elsewhere~\cite{detector}, 
only the components of the detector crucial to this analysis are
 summarized below. 
Charged particle tracking is provided by a five-layer silicon
vertex tracker (SVT) and a 40-layer drift chamber (DCH). 
For charged-particle identification, ionization energy loss ($dE/dx$) in
 the DCH and SVT, and Cherenkov radiation detected in a ring-imaging
 device (DIRC) are used. 
Photons and neutral pions are identified and measured using
the electromagnetic calorimeter, which comprises 6580 thallium-doped CsI
crystals. These systems are mounted inside a 1.5 T solenoidal
superconducting magnet. 
We use the GEANT~\cite{geant} software to simulate interactions of particles
traversing the \babar\ detector, taking into account the varying
detector conditions and beam backgrounds. 

\section{Analysis}

We select events with a minimum of three reconstructed charged tracks and
a total measured energy greater than 4.5 GeV, as determined
using all charged tracks and neutral clusters with energy above 30 MeV.
In order to reject $e^{+}e^{-}\to q\bar{q}, (q=u,d,s,c)$ continuum background, the ratio of the second
and zeroth order Fox-Wolfram moments~\cite{fox} 
must be less than 0.5.

The selection criteria are optimized
to maximize the ratio of the expected number of signal events 
over the square-root of the
expected number of background events, $S/\sqrt{B}$. The optimization was performed
on large samples of simulated signal and \qqbar\ and \BB\ 
background events.

The \Ds\ mesons 
are reconstructed in the modes
 $\Ds {\to} \phi \pi^+$, $\Kstarzb\Kp$ and $\Kzb (\KS) K^+$, 
with $\phi{\to} K^+K^-$, $\KS {\to} \pip \pim$ and $\Kstarzb{\to} K^-\pi^+$. 
The $\KS$ candidates are reconstructed from two
oppositely-charged tracks with an invariant mass 
$493 < M_{\pip \pim} < 503\mevcc $.  
All other charged tracks in the $B$ meson decay are required to 
originate from the $\epem$ interaction point.
Depending on the decay mode of the \Ds\ mesons, different selections
are used to identify charged kaons. In the $\dsphipi(\phi{\to}
K^+K^-)$ decay mode,  we identify kaons by applying a pion
veto with an efficiency of 95\% for kaons and a 20\% 
pion misidentification. In the \dskstark\ and \dsksk\ modes, a tight kaon
selection with an efficiency of 85\% and 5\% pion misidentification
probability is required for the $K^+$ candidate from the \Ds\ meson. 
%A
%loose selection with an efficiency of {\bf xx}\% and pion
%misidentification rate of {\bf yy}\% is applied to the $K^-$ candidates
%in the \kstarkpi\ decay chain. 

The $\phi$ candidates are reconstructed from two oppositely-charged 
kaons with an invariant mass $1009 < M_{\Kp \Km} < 1031\mevcc $. 
The $\Kstarzb$ candidates are constructed from the \Km\ and a \pip\
candidates and are required to have an invariant mass in the range
$862 < M_{\Km \pip} < 922~\mevcc $.

The polarizations of the 
 \Kstarzb\ and $\phi$ mesons in the  
\Ds\ decays are also utilized to reject backgrounds through the use of
the helicity angles
$\theta_H(\Kstarzb)$ and $\theta_H(\phi)$.  
The helicity angle is defined as the angle
between one of the decay products of 
the \Kstarzb\ (or $\phi$) 
and the direction of flight of the meson itself, in the meson rest
frame. 
Background events are distributed uniformly in $\cos\theta_H$ since they
originate from random combinations, while
signal events are distributed as $\cos^2\theta_H$.
Both $\phi$ and \Kstarzb\ candidates are required to have
$|\cos\theta_H|>0.55$. 
Finally, after constraining the \Ds\ decay products to the same
geometric vertex with a probability greater than $10^{-4}$, the \Ds 
candidates are required to have an invariant mass 
within 8 or 9 \mevcc\ of the known value~\cite{PDG2004}, 
depending on the \Ds\ mode.

%
% rho selection
%

The neutral pion candidates are reconstructed from a pair of photons
with a minimum energy of $30$~MeV. The invariant mass of the photon
pair is
required to be within a window $100 < m_{\gamma\gamma} < 160~\mevcc$.
After the mass of the $\pi^0$ candidate is constrained to 135~\mevcc,
it is combined with a track originating from the interaction point
to form $\rho^-$ candidates. The charged tracks are required to pass a
loose pion selection.
%%, which had the efficiency of {\bf xxx}\% and the
%%kaon misidentification probability of {\bf yyyy}\%. 
We require that
the invariant mass of the two 
pions forming the $\rho^-$ candidate be within 160~\mevcc\ of the known
value~\cite{PDG2004}. 

We also take advantage of the $\rho^-$
polarization in the $\btodsrho$ decays, requiring that the cosine of
the helicity angle $\theta_H(\rho)$ be either larger than $0.3$,
$0.35$, $0.2$, or smaller than $-0.15$, $-0.1$, $-0.4$ for the
\dsphipi, \dskstark, and \dsksk\ modes, respectively. Signal events
are distributed as $\cos^2\theta_H(\rho)$, modulated by the energy
dependence of the $\pi^0$ efficiency. The asymmetric selection takes into
account the larger probability to find a random combination of charged
and neutral pions in the forward direction in $\theta_H(\rho)$, which
corresponds to low $\pi^0$ energy. 

%
% B selection
%

We combine oppositely-charged $D_s^{\pm}$ and $\rho^\mp$ candidates
to form $\btodsrho$ candidates. The mass of $D_s^{\pm}$ candidates is
constrained to the known value~\cite{PDG2004}. 
In order to reject events where the \Ds\
comes from a $B$ and the $\rho^-$ from the other $B$, 
we require that
these two candidates have a vertex fit probability
greater than 0.6\% of originating from a common
vertex. 

We suppress combinatorial background from \qqbar\ production
using the event topology, computing the angle
($\theta_T$) between the thrust axis of 
 the $B$ meson decay product candidates and the thrust axis of all the other particles in the event.
In the center-of-mass frame (CM), \BB\ pairs are
produced approximately at rest and produce a uniform $\cos\theta_T$
distribution. 
In contrast,
\qqbar\ pairs are produced back-to-back in the CM frame,
which results in a $|\cos\theta_T|$
 distribution peaking at 1. 
Depending on the background level of each mode, $|\cos\theta_T|$ is
required to be smaller than a value which ranges between 0.6 and 0.85.

We also exploit the characteristic $\sin^2\theta_B$ angular distribution
for $e^+e^-\to B\overline{B}$ events, 
where $\theta_B$ is the angle between
the $B$ candidate flight direction and the direction of the 
incident electron beam
in CM frame. In contrast, the \qqbar\ backgrounds tend to
maintain $1+\cos^2\theta_B$ distribution characteristic of spin-$1/2$
particles. We require that $|\cos\theta_B|$ of the $B$ candidate 
is less than 0.6 for $\dsphipi$, 0.8 for $\dskstark$, and 0.7 for $\dsksk$.

\begin{figure}[h]
\begin{center}
\epsfig{file=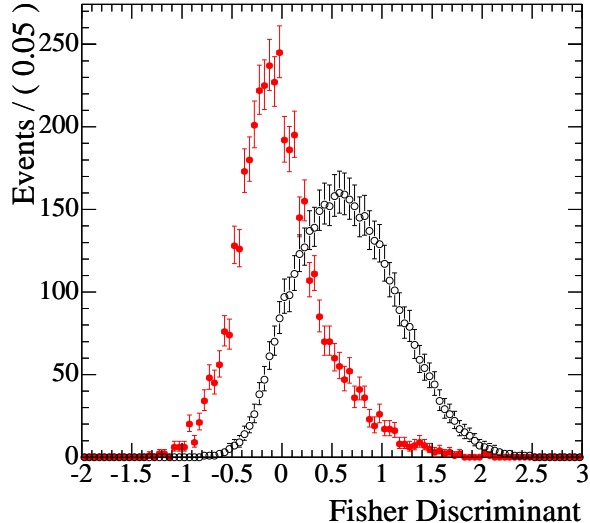,height=3in} 
\end{center}
\vspace{-1cm}
\caption{Fisher discriminant for simulated signal (solid red circles)
events and continuum background (open black circles) reconstructed as
\btodsrho,\dsphipi} 
\label{fig:fish} 
\end{figure}

 We further suppress backgrounds using a Fisher discriminant, $\cal{F}$, constructed from
the scalar sum of the CM momenta of all tracks and photons (excluding
the $B$ candidate 
decay products) flowing into 9 concentric cones centered on the thrust axis of the $B$ candidate~\cite{twobody}. The more spherical the
event, the lower the value of ${\cal{F}}$.   
Figure~\ref{fig:fish} shows a plot of the Fisher discriminant for signal and
continuum background events from Monte Carlo.

\begin{figure}[@tbp]
\begin{center}
\begin{tabular}{lr}
\epsfig{file=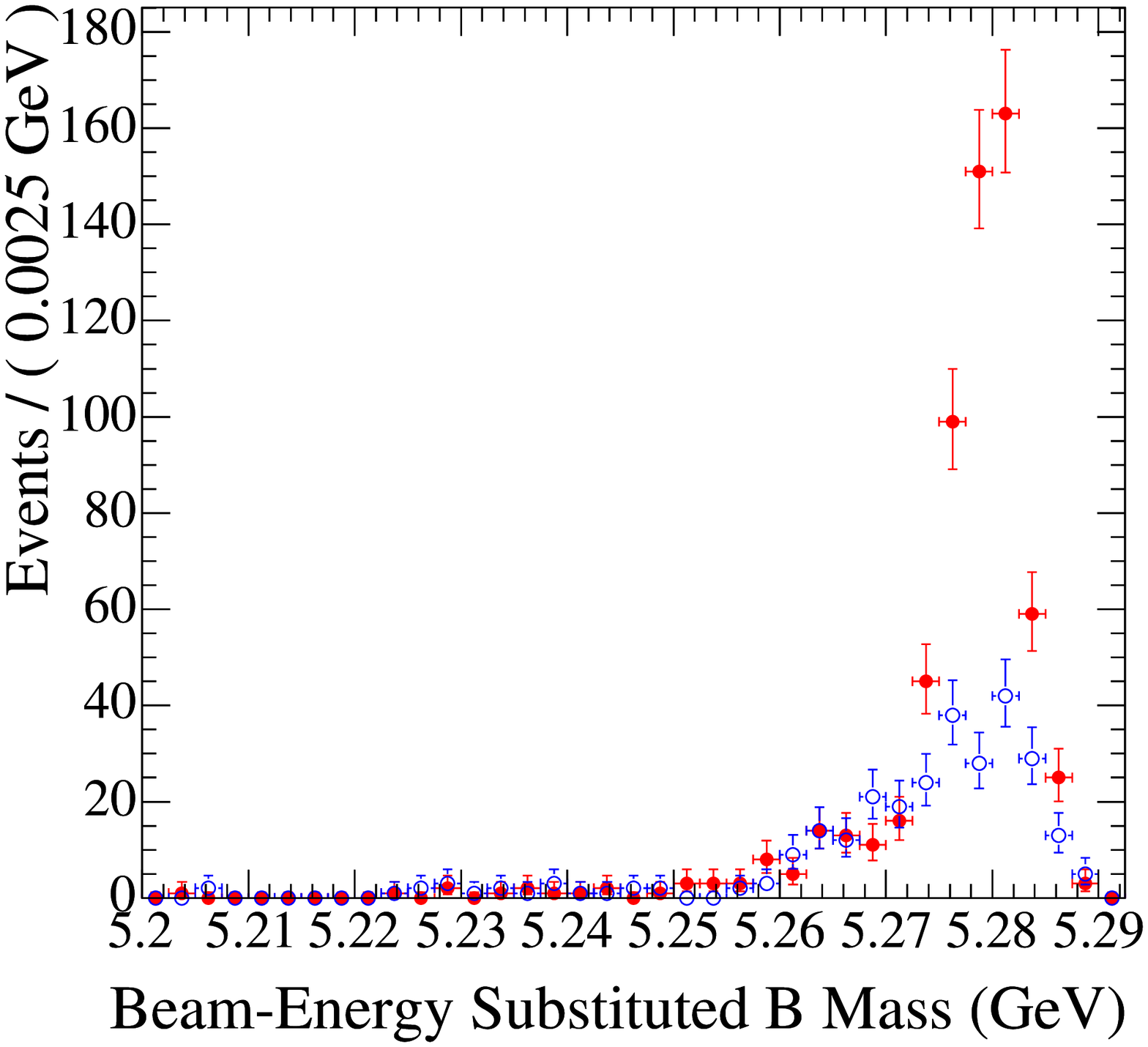,width=0.5\textwidth}& 
\epsfig{file=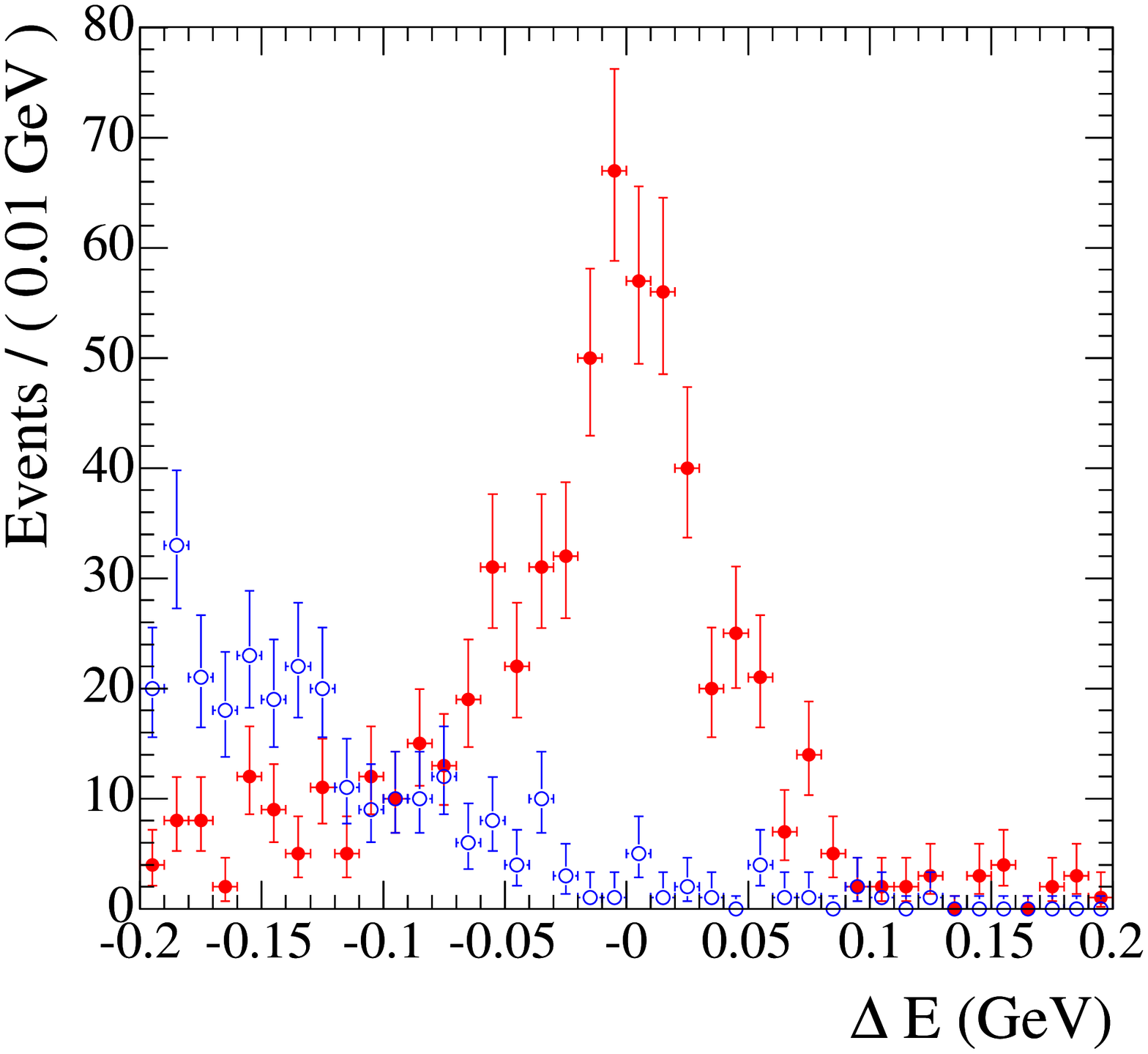,width=0.5\textwidth} \\
\end{tabular}
\vspace{-0.5cm}
\caption{\mse\ (left) and \De\ (right) distributions for simulated signal (solid red circles) and $B^0\to\Dss\rho^-$ (open blue circles) peaking background events reconstructed as $\btodsrho$, $\dsphipi$.}
\vspace{-0.5cm}
\label{fig:mesde} 
\end{center}
\end{figure}

We extract the signal using two kinematic variables \mes\ and
$\De$.  The first is the beam-energy-substituted mass 
$\mes = \sqrt{ (s/2 +
  \mathbf{p}_{i}\cdot\mathbf{p}_{B})^{2}/E_{i}^{2}- 
  \mathbf{p}^{2}_{B}}$, 
where $\sqrt{s}$ is the total \epem\ center-of-mass
energy,  $(E_{i},\mathbf{p}_{i})$ is the four-momentum of the initial
\epem\ system and $\mathbf{p}_{B}$ is the \Bz\ candidate momentum,
both measured in the laboratory frame. The second variable is 
$\De = E^{*}_{B} - \sqrt{s}/2$, where $E^{*}_{B}$ is the \Bz\ candidate
energy in the CM frame. 
For signal events, \mes\ peaks at the $B$ meson mass with 
a resolution of about 3 MeV$/c^2$ and \De\ peaks near zero
with a resolution of about 30~MeV, 
indicating that the candidate system of particles has a total energy
consistent with 
the beam energy in the CM frame. For the purposes of selection
optimization, we define the signal region between
$5.27<\mes<5.29$~GeV$/c^2$ and $|\Delta E|<45$~MeV. 
Figure~\ref{fig:mesde} shows the $\mes$ and $\Delta E$ distributions 
for simulated signal events and background from decays
$B^0\to\Dss\rho^-$ reconstructed as $\btodsrho$, $\dsphipi$.

Approximately 40\% of the selected events contained two or more
\btodsrho\ candidates that satisfy the criteria listed above. In
such events we select a single $B^0$ candidate based on:
1) the reconstructed mass of the \Ds\ meson, 2) the reconstructed mass of the
$\pi^0$ candidate, and 3) the \De\ variable.
The choice is made using these variables in a hierarchical manner 
with the \De\ selection
applied last, in order to avoid creating a bias in the \De\ distribution
of the background events. No significant bias is observed in the
large Monte Carlo sample of generic \qqbar\ production events and
$B^0$ and $B^\pm$ decays, excluding $b\to u\bar{c}s$ transitions. 

After the above selection, two main classes of backgrounds
remain.  First, there is combinatorial background from \qqbar
production and generic $B$ meson decays. We describe this background by a
two-dimensional distribution function of $\mes$ and
$\De$. In $\mes$, this background is 
described by an ARGUS function, $dN/dx\propto x\sqrt{1-2x^{2}/s}\exp\left[-\xi\left(1-2x^{2}/s\right)\right]$, 
characterized by the shape parameter $\xi$~\cite{argus}. 
In $\De$, the combinatorial background is well described by a
first-order polynomial. 

Second, $B$ meson decays with a $\Ds$ and a light meson in the final
state that correspond to $b\to u\bar{c}s$ quark transitions produce
distributions similar to signal events in $\mes$, but 
are typically shifted from zero or broadened in $\De$. These types of
background will, hereafter, be referred to as peaking backgrounds.
We have found 
that the largest contributions in the signal region come from decays
$B^+\to\Ds\rho^0$, $B^0\to\Dss\rho^-$, and $B^0\to\Dss\pi^-$. Of these
decay modes, only the latter has been previously
measured~\cite{dspi}. We determined the \mes\ and \De\ distributions
of these decays, 
reconstructed as signal $\btodsrho$ chain, from a large Monte Carlo
sample, equivalent to several times our data luminosity.  

Our kinematic selection suppresses higher
resonances such as $\rho(1450)$ and non-resonant $\Bz\to\Ds\pi^-\pi^0$
component, which 
are found to be negligibly small in $\Bz\to D^{-}\pi^+ \pi^0$
decays~\cite{ref:s2bgDRho}.
We ignore any possible contributions to $\Ds\pi^-\pi^0$ final state other
than $\rho^-\to\pi^- \pi^0$. 

%%% yield extraction

\section{Yield Extraction}

\begin{figure}
\begin{center}
\vspace{-1cm}
\epsfig{file=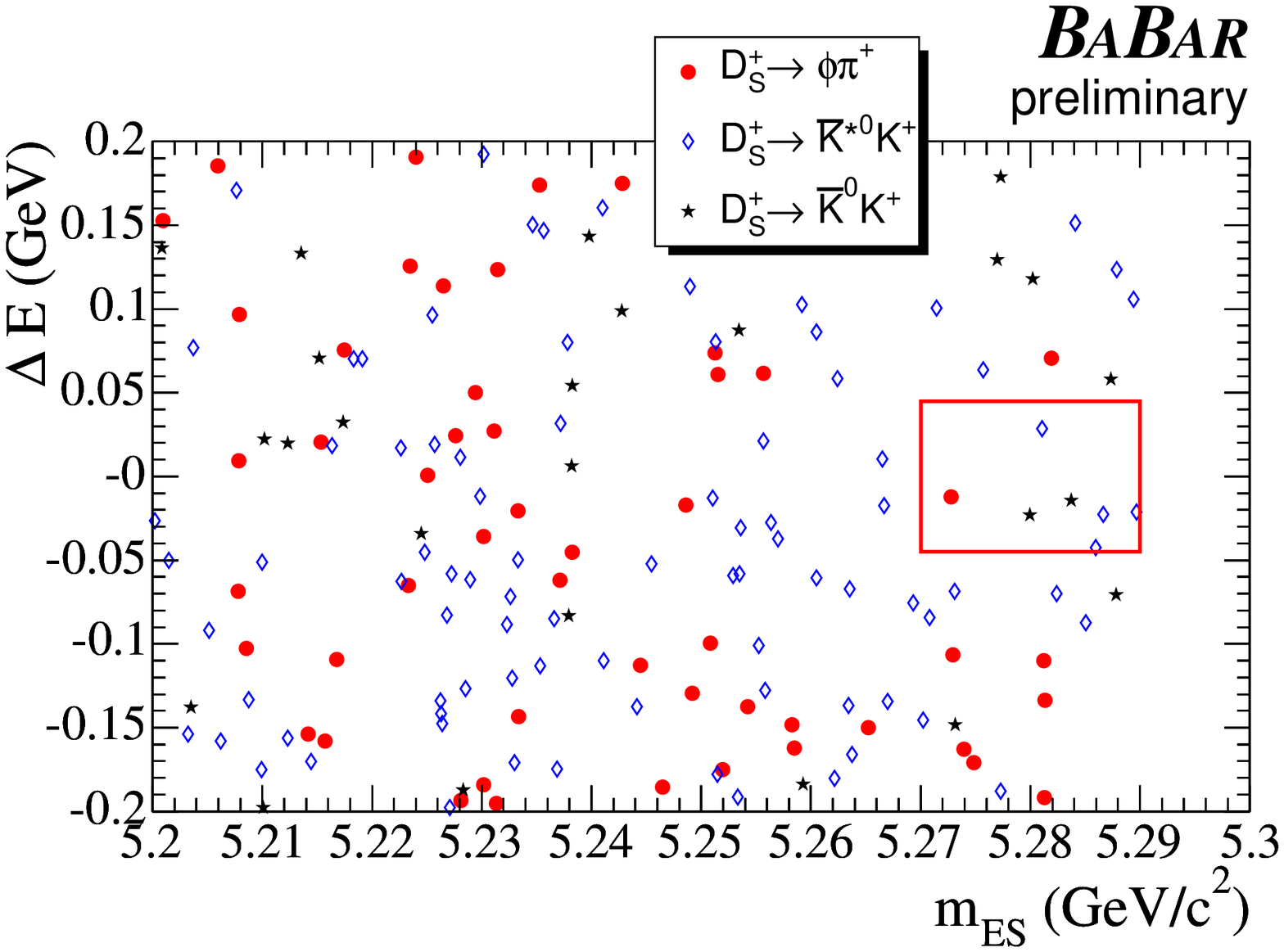,height=3in} 
\end{center}
\vspace{-0.8cm}
\caption{\De\ vs \mse\ distribution for \btodsrho\ candidates
reconstructed in the sample of \Lumi\ \FourS\ decays in \dsphipi\
(full circles), \dskstark\ (empty diamonds), and \dsksk\ (black stars)
modes. The box 
corresponds to the signal region $5.27<\mes<5.29~{\rm GeV}/c^2$ and
$|\De|<45$~MeV.}  
\label{fig:data} 
%\vspace{-0.5cm}
\end{figure}

Figure~\ref{fig:data} shows the distribution of events in the $(\mes,\De)$
plane for each of the \Ds\ decay modes. 
In the signal region,
we observe 1 event for \dsphipi, 4 for \dskstark, and 2 for \dsksk
, with comparable numbers expected from the
backgrounds. 

To extract the signal yield, we perform an unbinned extended
maximum-likelihood fit to the $(\mes,\De)$ distributions of all three
\Ds\ decay modes simultaneously in the ranges
$5.2\le \mes\le 5.3$~GeV/$c^2$, $-200\le\De\le 200$~MeV.  
The total number of events in the sample is 163. 
The likelihood function contains the contributions 
from the combinatorial backgrounds, signal, and peaking backgrounds. 
The probability density functions for combinatorial and peaking
backgrounds are found to be common to all three \Ds\ modes in Monte
Carlo simulation, while the shapes of the signal distributions are
determined independently for each \Ds\ decay mode. 
No significant correlation between \mse\ and \De\ is observed in Monte
Carlo samples,
and the likelihoods used in the fit ignore any such correlation. 

\begin{figure}
\begin{center}
\begin{tabular}{lr}
\epsfig{file=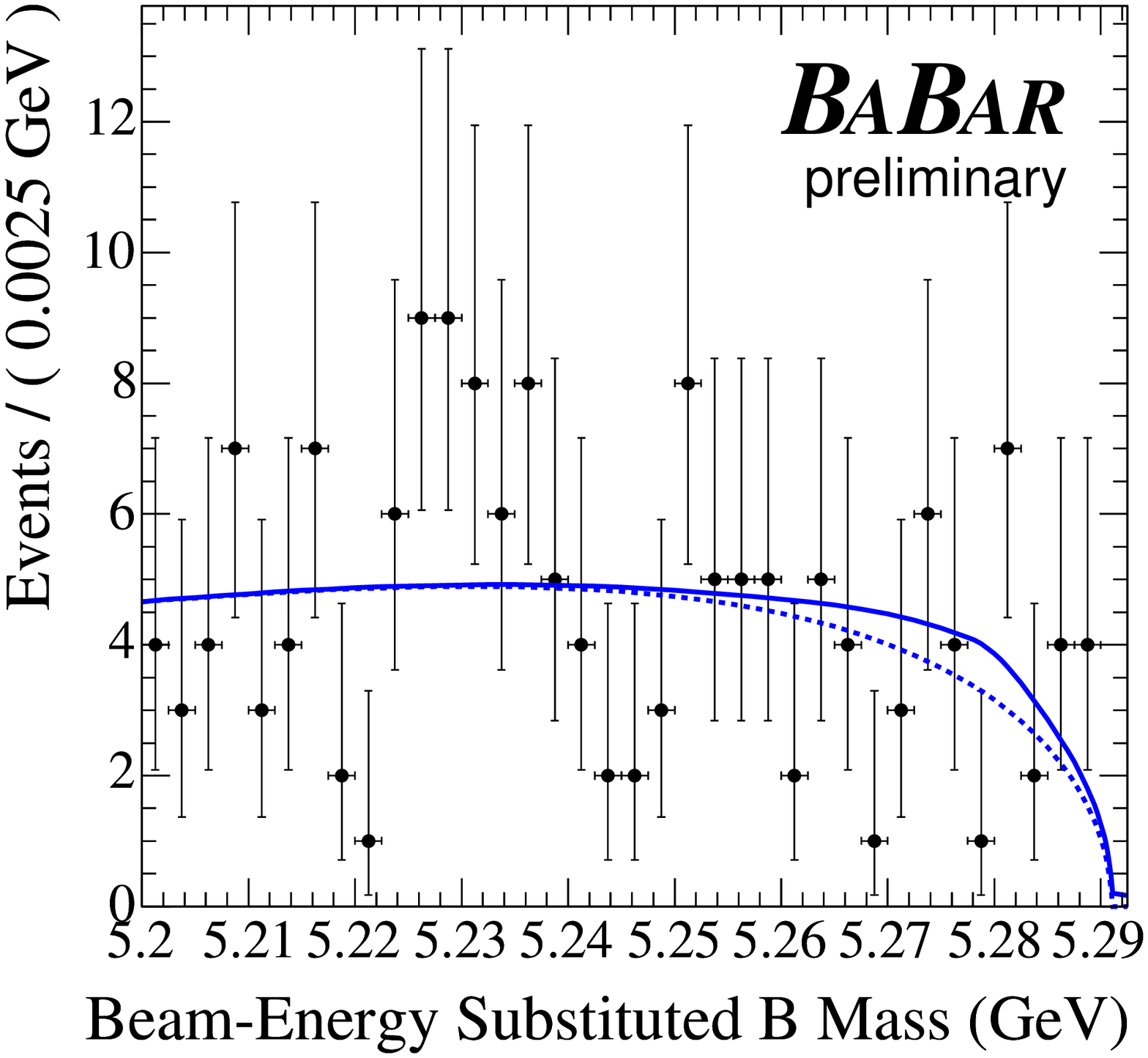,width=0.5\textwidth}& 
\epsfig{file=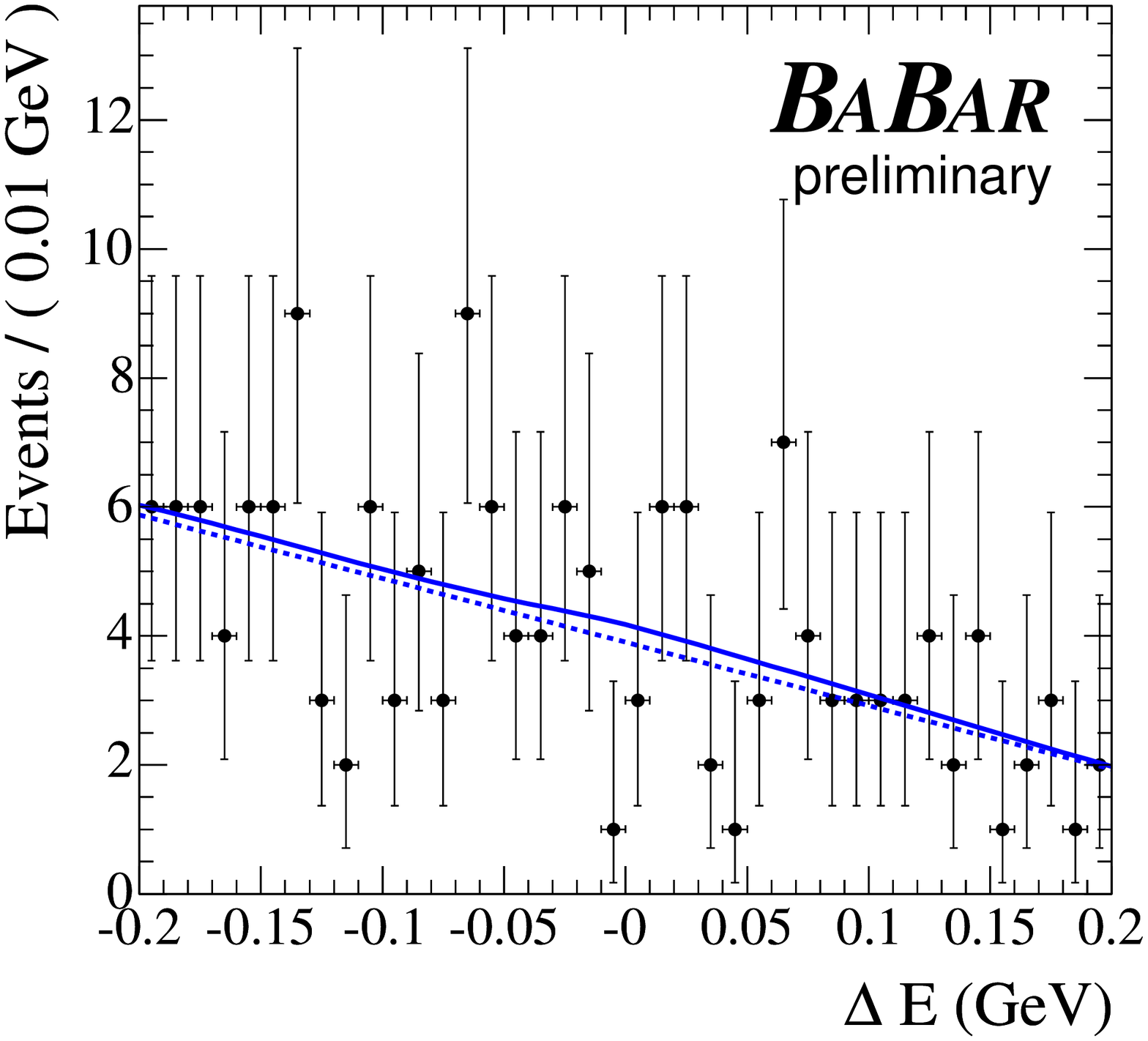,width=0.5\textwidth} \\
\end{tabular}
\caption{\mse\ (left) and \De\ (right) distributions for \btodsrho\ candidates
reconstructed in the sample of \Lumi\ \FourS\ decays. The solid curve
corresponds to the full probability density function from the combined
fit to all \Ds\ decay modes. The dashed line is the contribution from
the combinatorial backgrounds.}  
\label{fig:fit} 
\vspace{-0.5cm}
\end{center}
\end{figure}

Eight free parameters constrained by the fit include the shape of the
combinatorial 
background, characterized by the 
parameter $\xi$ in \mes\ and the constant term and a linear slope in
\De, as well 
as the combinatorial background yields in each \Ds\ mode and the
branching fractions of decays \btodsrho\ and $B^0\to\Dss\rho^-$. 
The signal and peaking background efficiencies are constrained to the
values determined by simulation for each \Ds\ decay mode. 
The branching fraction of
$B^+\to\Ds\rho^0$ is constrained to 
be half of $\BR(\btodsrho)$ from isospin symmetry, and the branching
fraction of $B^0\to\Dss\pi^-$, is fixed in the fit
to the value measured by \babar~\cite{dspi}. 
The uncertainties due to this assumption are included in the
systematics. 
The results of the fit
are shown in Fig.~\ref{fig:fit}. 

The reconstruction efficiencies, as estimated from
the Monte Carlo simulation, and the number of observed events in the signal region are summarized in the top portion of Table~\ref{tab:fit}. 
The bottom portion of Table~\ref{tab:fit} summarizes the signal, combinatorial background, and peaking background yields from the fit to data.
The central
value of the branching fraction returned by the fit is
$\BR(\btodsrho) = \left[0.2\pm 0.7 ({\rm stat.})\right]\times 10^{-5}$. 
The log-likelihood of the fit is consistent
with Monte Carlo expectations.

%% results
%
\begin{table}[@tbp]
\begin{center}
\caption{The reconstruction efficiency $\varepsilon$, number of
candidates $N_{\rm obs}$ observed in the signal region  
$5.27<\mes<5.29~{\rm GeV}/c^2$ and $|\De|<45$~MeV, the contributions
from the signal mode \btodsrho\ ($N_{\rm sig}$), and combinatorial
($N_{\rm comb}$) and peaking backgrounds from the fit to data.}
\label{tab:fit}
\vspace{0.2cm}
\begin{tabular}{|l|c|c|c||c|}
\hline
                  &          &           &        &                \\
                  & \dsphipi & \dskstark & \dsksk & All \ds\ Modes \\
\hline
$\varepsilon$ & $3.4\%$ & $1.4\%$ & $2.1\%$ & \\
$N_{\rm obs}$  & 1 & 4 & 2 & 7 \\
\hline
$N_{\rm sig}$     & $0.16 \pm 0.60$ & $0.06 \pm 0.21$ & $0.11 \pm 0.41$ &
   $0.32 \pm 0.75$ \\ 
$N_{\rm comb}$    & $1.64 \pm 0.26$ & $2.88 \pm 0.32$ & $0.74 \pm 0.18$ &
   $5.27 \pm 0.44$ \\
$N_{\btodssrho}$  & $0.04 \pm 0.14$ & $0.02 \pm 0.07$ & $0.03 \pm 0.10$ &
   $0.10 \pm 0.19$ \\
$N_{\btodsspi}$   & $0.42 \pm 0.29$ & $0.15 \pm 0.10$ & $0.30 \pm 0.21$ &
   $0.87 \pm 0.37$ \\ 
$N_{\bchtodsrho}$ & $0.01 \pm 0.01$ & $0.00 \pm 0.00$ & $0.01 \pm 0.01$ & 
   $0.02 \pm 0.01$ \\ 
\hline
\end{tabular}
\end{center}
\end{table}

\section{Systematic Uncertainties}

The systematic uncertainties are dominated by the potential contributions
from additional peaking backgrounds, such as decays \btodssk,
\bchtodssrho, \btodskstar, \btodsa, and \btodssa. Their contributions
are estimated from Monte Carlo. In addition, the branching fraction
of \btodsspi, fixed in the fit, is varied within its experimental
uncertainties~\cite{dspi}. Altogether, the uncertainties in the peaking
background contributions amount for a $\pm 0.2\times 10^{-5}$ systematic 
uncertainty on $\BR(\btodsrho)$. Other systematic uncertainties
include an 11\% relative uncertainty in 
$\BR(\Ds\rightarrow\phi\pip$)~\cite{phipi}, a 9\% relative uncertainty in
reconstruction efficiency due to Monte Carlo statistics, and
uncertainties in charged track reconstruction efficiency, $\pi^0$,
\KS, and charged kaon identification. 

%
% results
%
\section{Preliminary Results}

\begin{figure}[ht]
\begin{center}
\epsfig{file=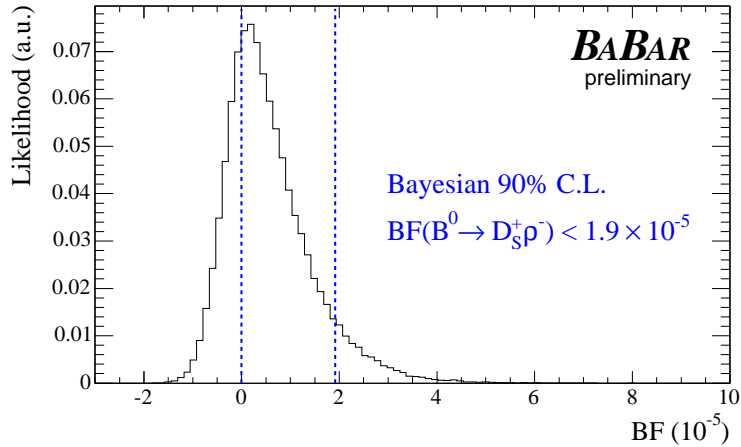,height=2.5in}
\end{center}
\vspace{-1cm}
\caption{Likelihood distribution for $\BR(\btodsrho)$, combining both
statistical and systematic uncertainties. The dashed lines show the
Bayesian 90\% confidence level limit assuming a constant prior for
$\BR(\btodsrho) > 0$.}
\label{fig:brLimit}
\end{figure}

Combining all systematic
uncertainties, the central value of \btodsrho\ branching fraction is 
\begin{displaymath}
\BR(\btodsrho) = \left[0.2\pm 0.7\ ({\rm stat.})\pm 0.2\ ({\rm syst.})\right]\times 10^{-5}\ ,
\end{displaymath}
consistent with zero within the current level of precision. 
We set a 90\% Bayesian confidence limit at
\begin{displaymath}
\brdsrholim ,
\end{displaymath}
assuming a constant prior for $\BR(\btodsrho) > 0$. The likelihood
distribution of $\BR(\btodsrho)$ and the 90\% C.L. limit are shown in
Fig.~\ref{fig:brLimit}. 

Using Eq.~(\ref{eq:rDRho}) and the recent lattice QCD value
$f_{D_s}/f_D=1.22\pm 0.04$~\cite{fdsd}, and assuming no additional
flavor SU(3) violation, we compute the value and 90\% confidence limit
on $r({D\rho})$ to be 
\begin{eqnarray}
r({D \rho}) &=& 0.003 \pm 0.006\ ({\rm stat.})\ \pm 0.002~({\rm syst.}) 
\nonumber\\
r({D\rho})&<&9.5\times10^{-3}\ ({\rm at~90\%~C.L.})
\nonumber
\end{eqnarray}

The low value of $r({D \rho})$ compared to $r({D^{(*)} \pi})$ determined
from \btodsospi\ decays~\cite{ref:s2bgDPi,dspi} is somewhat
unexpected. It implies small sensitivity of $C\! P$ asymmetries in
$\Bz{\to} D^{\mp}\rho^\pm$ decays to \stwobg, making that measurement
significantly more challenging.

% Standard acknowledgments paragraph; must always be included.
\section{Acknowledgments}

We are grateful for the 
extraordinary contributions of our \pep2\ colleagues in
achieving the excellent luminosity and machine conditions
that have made this work possible.
The success of this project also relies critically on the 
expertise and dedication of the computing organizations that 
support \babar.
The collaborating institutions wish to thank 
SLAC for its support and the kind hospitality extended to them. 
This work is supported by the
US Department of Energy
and National Science Foundation, the
Natural Sciences and Engineering Research Council (Canada),
Institute of High Energy Physics (China), the
Commissariat \`a l'Energie Atomique and
Institut National de Physique Nucl\'eaire et de Physique des Particules
(France), the
Bundesministerium f\"ur Bildung und Forschung and
Deutsche Forschungsgemeinschaft
(Germany), the
Istituto Nazionale di Fisica Nucleare (Italy),
the Foundation for Fundamental Research on Matter (The Netherlands),
the Research Council of Norway, the
Ministry of Science and Technology of the Russian Federation, and the
Particle Physics and Astronomy Research Council (United Kingdom). 
Individuals have received support from 
CONACyT (Mexico),
the A. P. Sloan Foundation, 
the Research Corporation,
and the Alexander von Humboldt Foundation.


\begin{thebibliography}{99}
\bibitem{chconj}
Charge conjugation is implied throughout this letter, unless
explicitly stated.
\end{thebibliography}
\end{document}